# High performance solar cells based on graphene/GaAs heterostructures

Xiaoqiang Li,[†,‡] Shengjiao Zhang,[†,‡] Peng Wang,[†,‡] Huikai Zhong,[†] Zhiqian Wu,[†] Hongsheng Chen,[†,‡] Erping Li,[†] Cheng Liu,[‡] Shisheng Lin[*,†,‡]

[†] Department of Information Science and Electronic Engineering, Zhejiang University, Hangzhou 310027, China

[‡] State Key Laboratory of Modern Optical Instrumentation, Zhejiang University, Hangzhou, 310027, China

[*] Corresponding author. Email: shishenglin@zju.edu.cn

**The honeycomb connection of carbon atoms by covalent bonds in a macroscopic two-dimensional scale leads to fascinating graphene and solar cells based on graphene/silicon Schottky diode have been widely studied. For solar cell applications, GaAs is superior to silicon as it has a direct band gap of 1.42 eV and its electron mobility is six times of that of silicon. However, graphene/GaAs solar cell has been rarely explored. Herein, we report graphene/GaAs solar cells with conversion efficiency (Eta) of 10.4% and 15.5% without and with anti-reflection layer on graphene, respectively. The Eta of 15.5% is higher than the state of art efficiency for graphene/Si system (14.5%). Furthermore, our calculation points out Eta of 25.8% can be reached by reasonably optimizing the open circuit voltage, junction ideality factor, resistance of graphene and metal/graphene contact. This research strongly support graphene/GaAs heterostructure solar cell have great potential for practical applications.**

As a two-dimensional (2D) giant flat molecule, graphene possesses a few outstanding electrical and optical properties, such as extremely high carrier mobility[1], micro-scale ballistic transport[2], abnormal quantum Hall effect[3], 2.3% constant absorption of visible light[4]. Besides, its low density of energy states near Dirac point and tunable doping concentration discriminate graphene from thin metals and traditional semiconductors[5]. Graphene also has extraordinary thermal conductivity[6] and high mechanical strength[7]. All these aspects make graphene a well-defined 'alien' in human-developed materials[1,8]. Attracted by those fascinating properties, graphene/semiconductor heterostructures are promising for solar cells applications[9-14]. Among them, graphene/silicon (Si) solar cell is most popular and much attention has been dedicated to it[15-19]. The state of art Eta of graphene/Si solar cell is limited to 8.6%[20] and 14.5%[15], without and with anti-reflection coating (ARC), respectively. Compared with Si, GaAs is commonly used to fabricate high efficient solar cells[21-24]. Suitable direct band gap energy of 1.42 eV and high electron mobility (8000cm$^2$/V s at 300K[25]), which is about six times of that of Si (1350 cm$^2$/V s at 300K[26]), make GaAs one of the best candidates for high performance solar cells[27]. Until now, there is rare work on graphene/GaAs solar cells[28]. Jie et al reported graphene/GaAs solar cells which can only convert 1.95% of input light into electricity[29], which is poor considering the advantages of GaAs over Si. Thus, a deep insight into graphene/GaAs solar cells is highly needed. Herein, we have achieved high performance solar cells with Eta of 10.4% for doped graphene/GaAs structure. Through anti-reflection technique, Eta has been further improved up to 15.5%, which is higher than the state of art Eta for graphene/Si system. It is noteworthy that 25.8% of Eta can be reasonably calculated for the van der Waals Schottky diode formed between graphene and GaAs, promising the practical application of graphene/GaAs heterostructure in solar cells.

## Structure of the graphene/GaAs solar cells

The schematic structure of the graphene/GaAs solar cell is illustrated in Fig. 1a, which is composed of GaAs substrate, graphene and electrodes. A $SiN_x$ film is sandwiched between graphene and GaAs as the dielectric insulating layer. The GaAs substrate is heavily *n*-type doped, which has a resistivity of 0.01-0.1 Ω cm. Nearly 50 graphene/GaAs solar cells have been fabricated in this work while the digital photograph of the typical graphene/GaAs solar cells is shown in Fig. 1b. The Raman spectrum of graphene is shown in Fig. 1c, where very weak defect-related D peak (around 1350 $cm^{-1}$) indicates the high quality of graphene. As seen from Fig. 1c, the G peak of as-grown graphene is located around 1596 $cm^{-1}$, which is blue-shifted compared with 1580 cm$^{-1}$ of undoped graphene, indicating as-grown graphene is p-type doped.

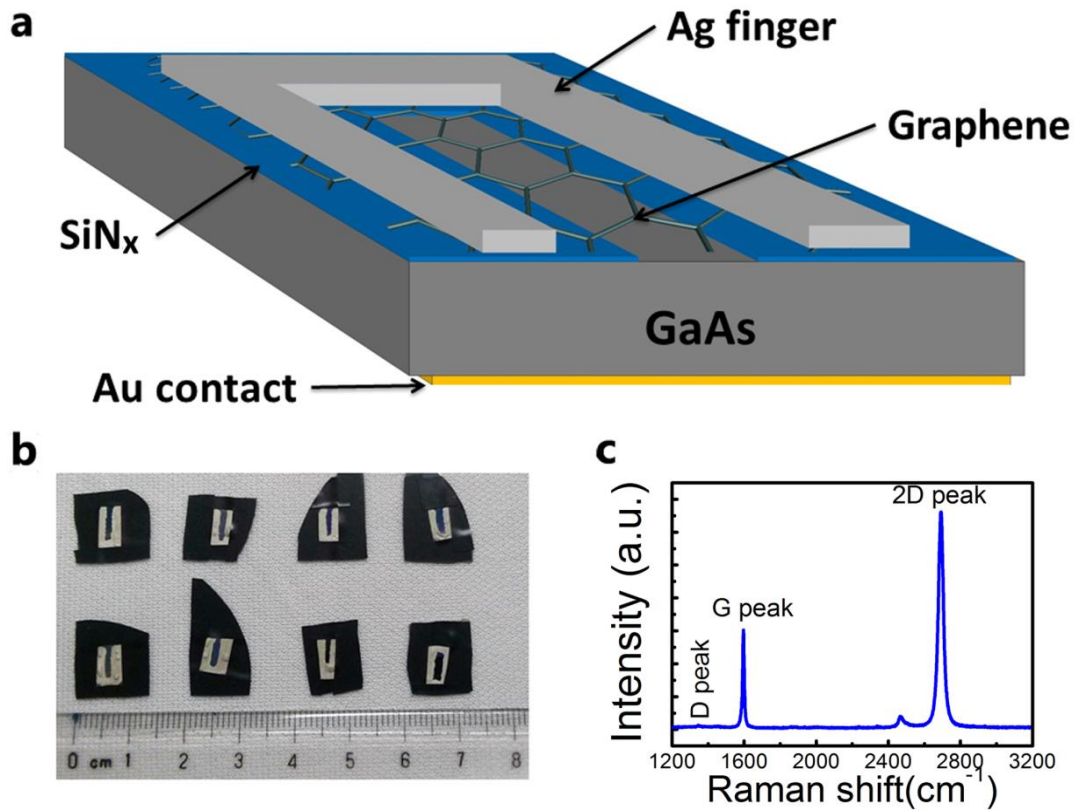

**Figure 1 | Graphene/GaAs solar cells. a**, Schematic structure of the graphene/GaAs solar cell. **b**, Photograph of graphene/GaAs solar cell samples. **c**, Raman spectrum of monolayer graphene in the devices.

## Depending of solar cell performance on graphene layer numbers

The schematic electronic band structure of graphene/GaAs Schottky diode is displayed in Fig. 2a. As the GaAs substrate is heavily *n*-type doped, its work function ($\Phi_{n\text{-GaAs}}$ in Fig. 2a) is in the range of 4.07±0.05eV, close to its electron affinity (4.07 eV in Ref. 28[30]). The electron affinity, i.e. energy difference between vaccum level and Dirac point of graphene is about 4.6eV[31]. As the Fermi level of graphene can be well adjusted by doping, its work function ($\Phi_{graphene}$ in Fig. 2a) is varied according to the doping concentration. When GaAs touches with graphene by van der Waals forces, Schottky junction is formed with a build-in barrier and a depletion region in GaAs. As the screening length of graphene is less than 0.5 nm[32] and GaAs is in bulk form, the traditional analysis based on bulk semiconductor physics is granted for graphene/GaAs heterostructure. The barrier height of this Schottky junction ($\Phi_{barrier}$ in Fig. 2a) is the difference between $\Phi_{n\text{-}GaAs}$ and $\Phi_{graphene}$. Dark current density-voltage (J-V) curve of the graphene/GaAs Schottky junction can be expressed by equation 1:

$$J = J_0 (\exp \frac{qV}{N_{IF} KT} - 1) \quad (1)$$

where $K$ is the Boltzmann constant, $N_{IF}$ is the junction ideality factor, q is the value of electron charge. Based on thermionic-emission theory, $J_0$ can be described as:

$$J_0 = AT^2 \exp(-\frac{q\Phi_{barrier}}{KT}) \quad (2)$$

where $A$ is the effective Richardson's constant of GaAs (8.9A/k cm$^2$) [33], $\Phi_{barrier}$ is the biuld-in junction barrier height. Under light illumination, the electrons and holes produced in GaAs substrate will be seperated by this Schottky junction and transport through GaAs and graphene, respectively.

The Eta of the Schottky junction depends on the short circuit current density ($J_{sc}$), open circuit voltage ($V_{oc}$) and fill factor (FF) based on the following equation: Eta=$V_{oc}$×$J_{sc}$×FF. Several papers have reported that graphene layer numbers affects the performance of this type solar cells[17,34,35]. In order to optimize graphene/GaAs solar cell, devices with different layer numbers of graphene are studied. The experimental value of $V_{oc}$ for the devices using as-grown graphene shows monotonically decreasing trend as the layer number increases, as shown in Fig. 2b. $V_{oc}$ of the graphene/GaAs Schottky diode is affected by carrier lifetime in bulk ($\tau_{bulk}$) and surface ($\tau_{surface}$), $\Phi_{barrier}$ and $N_{IF}$, which are independent in traditional solar cells[36]. It can be expected that increasing the layer numbers of graphene will decrease $J_{sc}$, which is induced by the enhanced light absorption of graphene layers. Monolayer graphene absorbs 2.3% of sun light[4], which is assumed to be dissipated and cannot be converted into electricity. Thus, $J_{sc}$ of the solar cells with multi-layers graphene is calculated through subtracting 2.3% from the experimental value of monolayer graphene/GaAs device while adding one more layer of graphene on the device. However, it is reasonable assumed that some of holes produced in GaAs are eliminated by defects or impurities while hopping among graphene layers. Thus, interlayer recombination is enhanced and $\tau_{surface}$ is reduced while stacking multilayer graphene over GaAs. Indeed, as shown in Fig. 2c, the experimental values of $J_{sc}$ are always lower than the calculated ones. On the other hand, $\tau_{bulk}$ is keeping constant for all the samples as it is decided by the GaAs substrate, while $\Phi_{barrier}$ and $N_{IF}$ vary according to layer numbers. $\Phi_{barrier}$ and $N_{IF}$ can be obtained through measured dark current-voltage (I-V) curve fitting by equation 1 and 2, which are shown in Fig. 2d. The values of $N_{IF}$ fall in the range of 1.7-2.1 and have no obvious relationship with graphene layer numbers. Similarly, as the graphene layer number increases, $\Phi_{barrier}$ exhibits a fluctuating behaviour, which cannot explain the decreasing trend of $V_{oc}$. Thus, it is most possibly decreasing values of $\tau_{surface}$ causes the decreasing $V_{oc}$ of the devices as graphene layer number increases, which aggrees with the assumption of interlayer recombinations. It is noteworthy trilayer graphene based solar cell has the lowest $N_{IF}$ and highest $\Phi_{barrier}$.

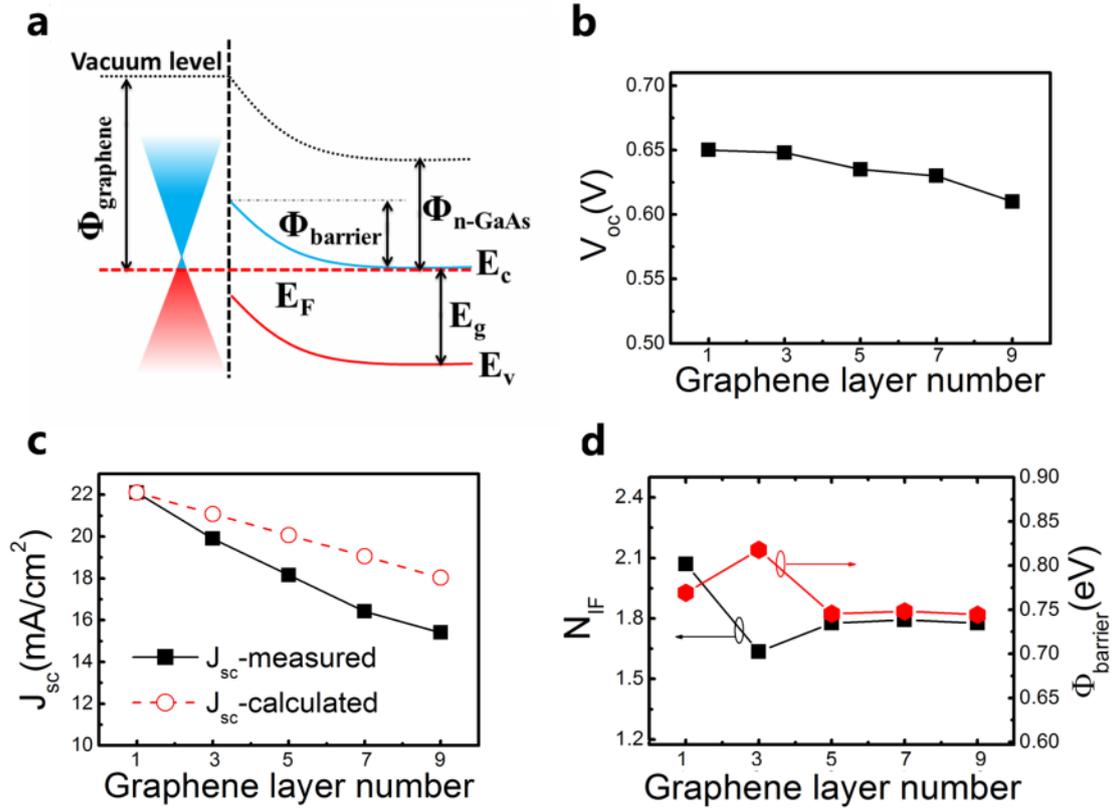

**Figure 2 | Band structure of graphene/GaAs Schottky junction and the effect of graphene layer numbers on physical parameters of graphene/GaAs solar cells. a**, Band structure of graphene/GaAs Schottky junction. **b**, Measured $V_{oc}$ of the solar cells with different layers of graphene. **c**, Calculated and measured $J_{sc}$ with different layers of graphene. **d**, Deduced $N_{IF}$ and $\Phi_{barrier}$ of the solar cells with different layers of graphene.

As graphene layer number increases, FF value of the graphene/GaAs solar cell gradually increases as shown in Fig. 3a, which also agrees well with the calculated FF data. The calculation program of FF is based on classic solar cell FF model[37] (see Fig. S1). Briefly, FF is influenced by the equivalent serial connected resistance to the solar cell defined as series resistance ($R_s$), the equivalent parallel resistance causing leakage current defined as shunt resistance ($R_{shunt}$) and $N_{IF}$, which are calculated based on theoretically fitting of measured I-V data (Supplementary Fig. S1, S3 and S4). Fig. 3b shows the $R_s$ and $R_{shunt}$ values for the devices with different layers of graphene. As seen from Fig. 3b, $R_s$ decreases as the graphene layer number increases while $R_{shunt}$ scatters ranging from $1.15 \times 10^4 \, \Omega$ to $1.40 \times 10^4 \, \Omega$. The scattering of $R_{shunt}$ in this range can only induce 0.3% change of FF (Supplementary Fig. S1). Moreover, the scattering of $N_{IF}$ in a small range also has little effect on FF. Thus, changes of FF values of solar cells with different layers of graphene are mainly caused by variation of $R_s$.

Fig. 3c shows the experimental Eta results, where trilayer graphene based device has the highest Eta. Fig. 3c also displays the calculation results of Eta as a function of graphene layer number while assuming a constant $V_{oc}$ value of 0.65V for all the devices. It is also shown trilayer graphene based solar cell has the highest Eta, as the performance of solar cell with monolayer graphene is limited by low FF while the decreased $J_{sc}$ and $V_{oc}$ degrade Eta when graphene layer number becomes larger than three. Here it should be noted that the measured square resistance values ($R_{sq}$) of as-grown graphene are about 2000 $\Omega$/sq. For the trilayer graphene/GaAs solar cell with 8.9% of Eta, the reason why FF deviates from ideal value is analyzed. As the three limiting factors, $R_{shunt}$ of this device equals to $1.35 \times 10^4 \Omega$ while $N_{IF}$ and $R_s$ are 1.92 and $24.3\Omega$, respectively. As shown in Fig. 3d, for the deviation from ideal FF value, $N_{IF}$ contributes 51.8% while $R_s$ and $R_{shunt}$ contributes 46.3% and 1.9%, respectively. Responsible for almost half of the FF drop, $R_s$ is studied in detail for further optimization. Fig. 3e shows the schematic physical illustration of detailed constitution of $R_s$, which includes $R_{sq}$ of graphene, GaAs substrate resistivity ($R_{sub}$) and front Ag finger line resistivity ($R_{Ag}$), contact resistance between Ag paste and graphene ($R_{f-contact}$), contact resistance between Au and GaAs substrate ($R_{r-contact}$). Fig. 3f charts the proportion of each constituent part of $R_s$. It can be seen 50.9%, 33.1% and 15.5% of $R_s$ comes from $R_{f-contact}$, $R_{sq}$ and $R_{r-contact}$, respectively. On the other hand, $R_{sub}$ and $R_{Ag}$ have little effect on $R_s$. As the $R_{sq}$ of graphene can be further reduced down to $100\ \Omega/sq$ and $R_{f-contact}$ can be further reduced through choosing metal for better ohmic contact[38], improvements of FF can be strongly granted. Thus, doping of grapheme is introduced to reduce the $R_{sq}$ of graphene and optimize FF and Eta as below.

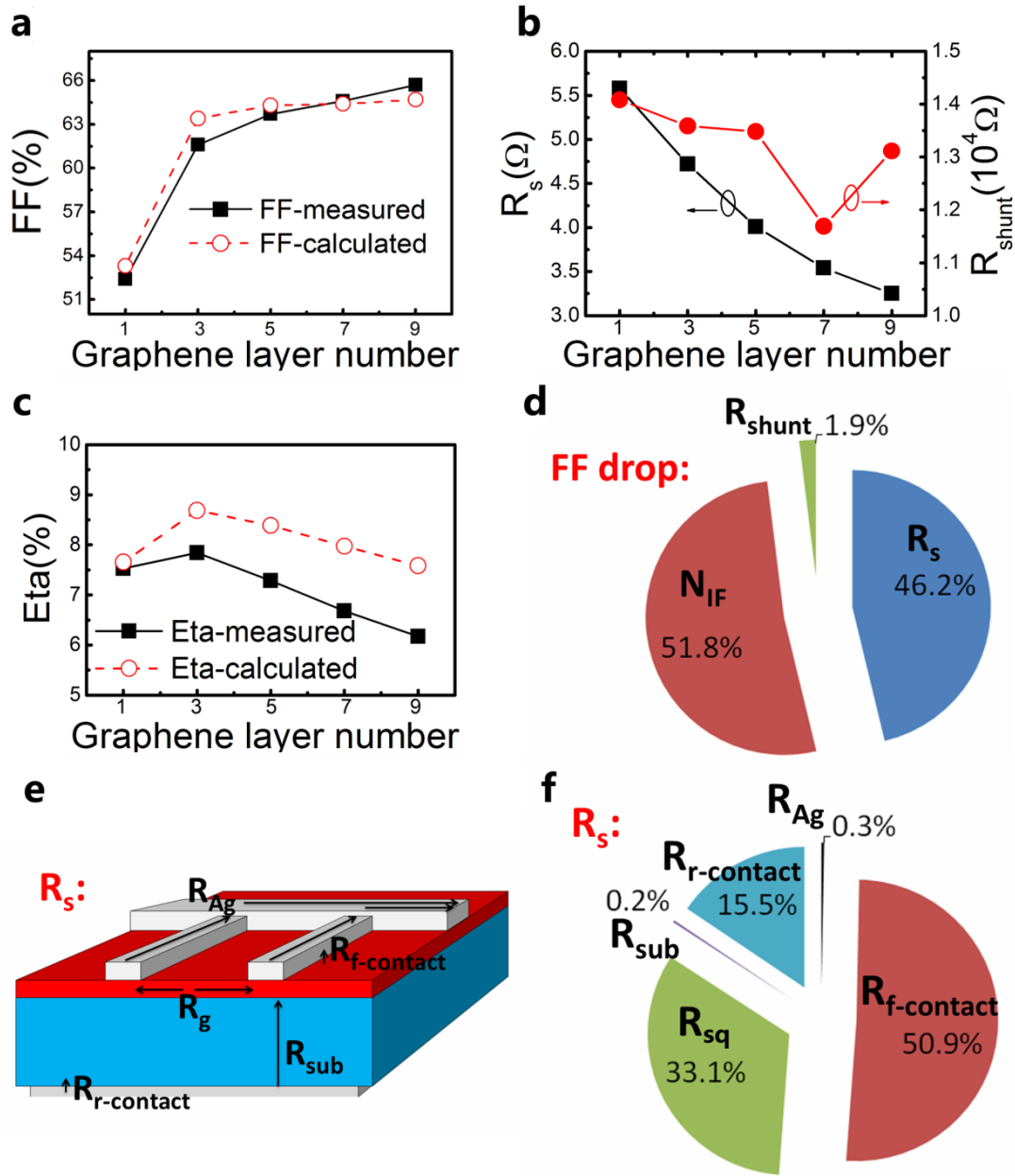

**Figure 3 | Effect of graphene layer number on FF and Eta of graphene/GaAs solar cells. a**, Calculated and measured FF of the solar cells with different layers of graphene. **b**, $R_s$ and $R_{shunt}$ of the solar cells with different layers of graphene. **c**, Calculated and measured Eta of the solar cells with different layers of graphene. The $R_{sq}$ of graphene is about 2000 Ω/sq. **d**, Proportion of FF drop induced by $R_s$, $R_{shunt}$ and $N_{IF}$. **e**, Schematic diagraph of the $R_s$ compositions. **f**, Proportion of each

constituent part of $R_s$.

## Optimize graphene/GaAs solar cells through doping and ARC

It is stressed $R_s$ and $N_{IF}$ can be optimized in the following experiments, which can improve FF[39]. As a demonstration, we can improve $R_s$ and $N_{IF}$ through doping graphene with bis(trifluoromethanesulfonyl)amide [(($CF_3SO_2$)$_2$NH)] (TFSA). By shifting the Fermi level of graphene under doping, we can also improve $V_{oc}$. We have measured several as-grown trilayer graphene/GaAs solar cells and it is found that a best Eta of 9.2% can be achieved with active area of 9.6mm$^2$, which is shown in Fig. 4a. Fig. 4a also shows the J-V curve recorded from the trilayer graphene/GaAs solar cells with TFSA doping and TFSA doping & ARC. For the device without doping, $V_{oc}$, $J_{sc}$ and FF are 0.72V, 22.4mA/cm$^2$ and 56.9%, respectively. Through TFSA doping, $R_{sq}$ of graphene is decreased from about 2000Ω/sq to 1000 Ω/sq. The G peak position in Raman spectrum for the as grown and TFSA doped graphene is 1596cm$^{-1}$ and 1601cm$^{-1}$ respectively (Supplementary Fig. S5). Accordingly, the Fermi level of as-grown and doped graphene is 0.38eV and 0.50eV below Dirac point respectively[40]. After doping, FF is improved from 56.9% to 62.8%. Meanwhile $V_{oc}$ increases to 0.76V as $\Phi_{barrier}$ increases, while $J_{sc}$ slightly decreases to 21.9 mA/cm$^2$. The best Eta of the TFSA doped graphene/GaAs solar cell is 10.4%.

Further improvements can be realized by improving $J_{sc}$ through introducing ARC layer on graphene. For the graphene/GaAs solar cells with ARC, $J_{sc}$ can be increased by about 30% compared with that of the devices without ARC, as well as slightly $V_{oc}$ increase, which may be attributed to surface passivation effect of the ARC layer. Eta of the best device with ARC is 15.5%, which is higher than the state of art Eta of 14.5% for graphene/Si solar cells[15]. Even though, the measured FF (69.9%) is still low compared with the thin film GaAs solar cells (typically above 80%)[41]. As mentioned above, further FF improvement can be achieved given low $N_{IF}$ and $R_{f-contact}$. Furthermore, it has been reported that Fermi level of graphene can be tuned as large as about 1.0eV below Dirac point[42], which can rise $\Phi_{barrier}$ of graphene/GaAs Schottky junction up to 1.5eV. While in our work TFSA doping only increase $V_{oc}$ by 0.04V, suggesting there is still large room for the improvement of $V_{oc}$. Considering $R_{f-contact}$ down to $10^{-5}$ Ω as reported[43], $N_{IF}$ value of the junction down to 1.2[44] and $V_{oc}$ value of 1.1V, Eta of graphene/GaAs solar cell with ARC can reach as high as 25.8% theoretically as shown in Fig. 4b, which is quite promising to be realized in the near future.

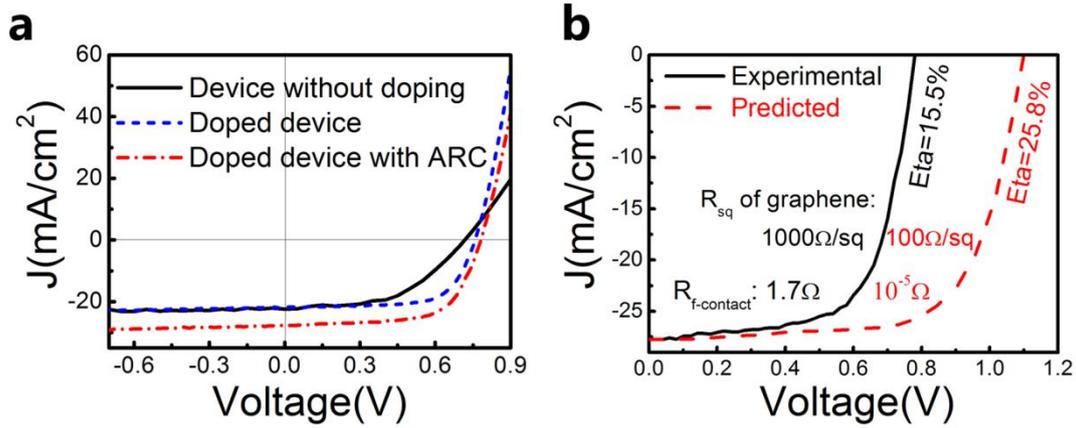

**Figure 4 | Experimental and calculation predicted J-V curves of graphene/GaAs solar cells. a**, Experimental $J-V$ curves of graphene/GaAs solar cells with as-grown graphene, with doping and with doping & ARC. **b**, Experimental and calculation predicted J-V curves of the best graphene/GaAs solar cells.

Fig. 5a shows the wavelength dependent external quantum efficiencies (EQE) and Fig. 5b shows the wavelength dependent reflectance of the best doped graphene/GaAs solar cells with and without ARC. As seen from Fig. 5a, EQE of the whole effective light spectrum (300nm-873nm) is in the range of 40% to 60% for the device without ARC, while it is improved to the range of 65% to 85% for the device with ARC. Fig. 5b shows the reflectance decreases from above 30% down to below 10% after ARC, which induces the increase of the EQE as shown in Fig.5a. As a result, $J_{sc}$ of the solar cell is increased by about 30%. As also can be seen in Fig. 5a, EQE falls into zero when the photon energy is lower than 1.42 eV of GaAs band gap (873nm in wavelength), which clarifies that sun light absorbed in graphene cannot be converted into electricity and demonstrates that only the electrons and holes produced in GaAs substrate is active for the electricity conversion in the device.

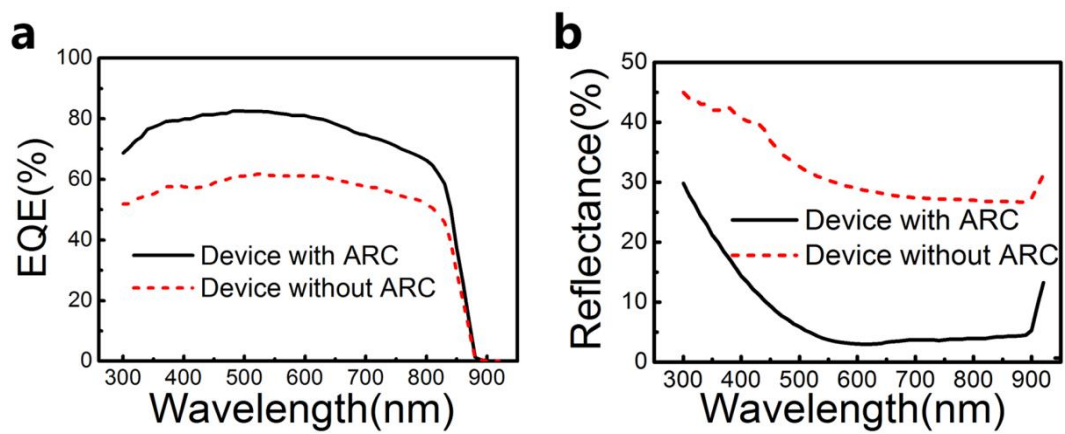

**Figure 5 | Spectral distribution of the EQE and reflectance of the graphene/GaAs solar cells. a**, EQE of the best device with and without ARC. **b**, Reflectance of the best device with and without ARC.

## Conclusions

High conversion efficiency of 15.5% has been achieved for $Al_2O_3$ coated graphene/GaAs solar cell. Combing experimental and theoritical work, we have looked insight into the factors that influencing $V_{oc}$, $J_{sc}$ and FF. The main limiting factors including $N_{IF}$ and $R_s$ are investigated in detail. Moreover, further improvements of the graphene/GaAs solar cell is predicted, pointing out that 25.8% conversion efficiency can be obtained with reasonably optimized junction quality, resistance of graphene sheet and metal/graphene contact. Besides, even higher conversion efficiency can be expected when the doping concentration of GaAs substrate is optimized. It is noteworthy that the bonds between graphene and GaAs are van der Waals forces and thus the formed Schottky diode is different from the bulk ones. Thus, further insight into this type of Schottky diode is highly desired in the following works to improve the value of $\tau_{surface}$, thus decreasing $N_{IF}$[45,46]. This work demonstrates that graphene/GaAs Schottky diode is a very promising structure for high efficient and practical useful graphene solar cells.

## Methods

**Solar cell samples fabrication.** Monolayer graphene was grown on copper substrate by chemical vapor deposition (CVD) technique using $CH_4$ and $H_2$ as the reaction source [47]. The growth was carried out at 1000℃ for 60min with reaction source flux ratio of $CH_4$:$H_2$ equals to 5:1. 60nm Au was thermally evaporated on GaAs for achieving Ohmic contacts. 80nm $SiN_x$ was deposited using plasma enhanced CVD and used as the dielectric layer under the graphene contact, and the opened window on $SiN_x$ film defined the solar cell active area. Graphene was transferred onto single side polished GaAs (001) substrate using PMMA as the sacrificing layer. Prior to transferring graphene onto GaAs surface, the opened area was cleaned by dipping the samples into 10% HCl solution for 5min followed with DI water rinse. After graphene transferring, silver was pasted onto graphene above $SiN_x$ coated area, followed by 120°C/5min post anneal. TSFA spin coating method was used for p-type doping of graphene[20]. Anti-reflection layer was realized by an $Al_2O_3$ film deposited by electron beam evaporation technique.

**Solar cells characterization.** The graphene/GaAs solar devices were tested with a solar simulator under AM1.5 condition. The current-voltage data were recorded using a Keithley 4200 system. The monolayer graphene was also transferred to $Si/SiO_2$ substrate and characterized by Raman spectroscopy (Renishaw inVia Reflex) with the excitation wavelength of 532nm. Non-active area was covered by black tape to minimize the test error. External quantum efficiencies of the best graphene/GaAs solar cells with and without ARC were measured with PV Measurements QEXL system. $N_{IF}$ was also tested with Suns-Voc measurement, which was carried out in the platform of Sinton WCT-120.

## Acknowledgments

S. S. Lin thanks the support from the National Science Foundation of China (No. 51202216, 61431014 and 61171037) and X. Q. Li thanks the support from the Postdoctoral Science Foundation of China (111400-X91305).

## Author contributions

S. S. Lin initiated this work. X. Q. Li, S. S. Lin, S. J. Zhang, P. Wang, H. K. Zhong and Z. Q. Wu performed the experiments. X. Q. Li, S. S. Lin, H. S. Chen and C. Liu analyzed the data. S. S. Lin designed and directed the work. S. S. Lin and X. Q. Li write the paper and all authors comment on this paper.

## Competing financial interests

The authors declare no competing financial interests.